\documentclass{edm_article}

\usepackage{cleveref}
\usepackage{xcolor}
\usepackage[utf8]{inputenc}
\usepackage[T1]{fontenc}
\usepackage[citestyle=numeric,bibstyle=numeric,backend=bibtex,eprint=false,doi=false,isbn=false,url=false]{biblatex}
\usepackage{multirow}
\usepackage{array}
\usepackage{tikz}
\newcommand{\PreserveBackslash}[1]{\let\temp=\\#1\let\\=\temp}
\newcolumntype{C}[1]{>{\PreserveBackslash\centering}p{#1}}
\newcolumntype{R}[1]{>{\PreserveBackslash\raggedleft}p{#1}}
\newcolumntype{L}[1]{>{\PreserveBackslash\raggedright}p{#1}}

\begin{document}

\title{Analyzing Student Strategies In Blended Courses Using Clickstream Data}

\numberofauthors{3}
 \author{
 \alignauthor
 Nil-Jana Akpinar\\
        \affaddr{Carnegie Mellon University}\\
        \affaddr{5000 Forbes Avenue}\\
        \affaddr{Pittsburgh, PA 15213}\\
        \email{nakpinar@stat.cmu.edu}
 \alignauthor
 Aaditya Ramdas\\
        \affaddr{Carnegie Mellon University}\\
		\affaddr{5000 Forbes Avenue}\\
		\affaddr{Pittsburgh, PA 15213}\\
        \email{aramdas@stat.cmu.edu}
 \alignauthor
 Umut Acar\\
        \affaddr{Carnegie Mellon University}\\
		\affaddr{5000 Forbes Avenue}\\
		\affaddr{Pittsburgh, PA 15213}\\
        \email{umut@cs.cmu.edu}
    }

\maketitle

\begin{abstract}
Educational software data promises unique insights into students' study behaviors and drivers of success.
While much work has been dedicated to performance prediction in massive open online courses, it is unclear if the same methods can be applied to blended courses and a deeper understanding of student strategies is often missing. 
We use pattern mining and models borrowed from Natural Language Processing (NLP) to understand student interactions and extract frequent strategies from a blended college course.
Fine-grained clickstream data is collected through Diderot, a non-commercial educational support system that spans a wide range of functionalities.
We find that interaction patterns differ considerably based on the assessment type students are preparing for, and many of the extracted features can be used for reliable performance prediction.
Our results suggest that the proposed hybrid NLP methods can provide valuable insights even in the low-data setting of blended courses given enough data granularity.

\end{abstract}


\section{Introduction}
\label{sec:introdcution}
Data collected through educational software systems can provide promising starting points to address hard questions rooted in the learning sciences.  %
Modern education relies increasingly on these systems to assist teaching and grading, manage learning content, provide discussion boards, facilitate group work, or replace the traditional class room setting altogether.
While blended courses revolve around the traditional class room setting accompanied by task-specific software support, Massive Open Online Courses (MOOCs) are usually entirely virtual and often involve video lectures and hundreds to thousands of students in a single course.
Almost by design, these systems come with unprecedented opportunities for large scale data collection on students' study habits, content exposure and learning trajectories. 

Much of the previous research effort has been directed towards performance prediction with the overall rationale that reliable estimation of students' grades and dropout probability at early course stages can be used to devise Early Warning Systems (EWSs) \cite[e.g.][]{macfadyen_mining_2010, hu_developing_2014, liz-dominguez_predictors_2019, bakhshinategh_educational_2018}. %
Despite considerable success in this area, many performance prediction models suffer from a list of shortcomings.
Prior work on performance prediction from student online activity data has predominantly focused on MOOCs \cite[e.g.][]{brooks_time_2015, li_exploring_2017,kloft_predicting_2014}, and it is unclear if the same methods can be applied to blended courses \cite{krauss_can_nodate}. In most blended courses, some of the learning activity takes place offline and cannot be tracked which leads to relatively shallow data on only fragments of courses.
In addition, many of the features that can be derived are simple and coarse summary statistics of students' online activity data, e.g. counts of clicks or logins, that only have a limited capacity to reflect the often complex strategies students take when interacting with course material.

A detailed understanding of how students interact with educational systems and the strategies they take is crucial for reliable performance prediction. We thus seek to understand how students approach learning in blended courses based on the second half of a sophomore level college course in computer science.
Our data is drawn from Diderot, a non-commercial educational software system developed at Carnegie Mellon University which spans functions for virtually all course components outside of face-to-face class and recitation times, and thereby allows us to overcome many of the challenges that are generally faced when mining blended courses. %
Despite evident similarities, there are several important characteristics which differentiate our blended learning setting from the study of MOOCs.
Most importantly, our data spans relatively few students and student actions which constitutes a challenge for many of the previously proposed methods. In addition, we have access to data that is unique to in-person classes such as individual attendance, and 
the nature of our activity data facilitates contextualization of student behavior which promises to increase the interpretability of downstream prediction models.

In this paper, we place a dual focus on methodology and educational insights.
On the one hand, we propose new modeling pipelines based on ideas from natural language processing that work well in the low-data setting of blended courses. 
On the other hand, we apply both new and existing methods to Diderot data and gain valuable insights into student behavior while addressing the following research questions:

\vspace{-2ex}
\begin{itemize}
	\item [\textbf{RQ1}] How do students interact with course material, and what are frequent strategies they take?
	\item [\textbf{RQ2}] How do students use these strategies for homework solving as compared to exam preparation?
	\item [\textbf{RQ3}] Are student strategies indicative of grade outcomes?
\end{itemize}
\vspace{-2ex}

The remainder of this paper is outlined as follows.
We discuss related work in \Cref{sec:relatedwork}, and proceed to give some context for the data in \Cref{sec:data}. \Cref{sec:methods} describes our methods including the preprocessing of clickstream data, and we discuss our results in \Cref{sec:results}. Finally, conclusions are drawn in \Cref{sec:conclusion}.

\section{Related Work}
\label{sec:relatedwork}

\subsection{Analysis Of Online Student Behavior}
Raw data from educational software systems often comes in the form of time-stamped student actions with an array of suitable identifiers.
Evidence for correlations between activity log-based features and performance outcomes are plentiful.
Many of the commonly discussed features revolve around simple summary statistics such as counts of certain types of actions, and have been shown to be indicative of students' success particularly in MOOCs.
Recent lines of research find links between general course completion in MOOCs and the number of watched videos \cite{pursel_understanding_2016, chen_mooc_2017}, the number of question answer attempts \cite{chen_mooc_2017}, and the time spent on assignments \cite{andres_studying_2018}.
Similar results have been observed for blended courses but are much scarcer \cite{sheshadri_predicting_2018,gitinabard_what_2019}.
In \cite{gitinabard_what_2019}, the authors analyze sequences of transitions between different online platforms in two undergraduate level college courses. Their study finds that, although students are generally more likely to stay on the same platform in a study session, high achieving students transition more often and are more likely to use the discussion board.
In many cases, the limited amount of data in blended courses is problematic and can lead to complications such as zero-inflated count variables.

A major shortcoming of count-based methods is their failure to leverage the sequential structure of students' interactions with educational software systems.
Both the order and the time difference between actions promise to carry valuable information that can be taken into consideration when relying on sequence based methods instead.
In this work, we propose a pipeline for analyzing student online behavior based on session study sequences. While the order of actions is taken into account explicitly, time differences help us to derive reliable study sessions.

\subsection{Study Sessions}

Sequence-based approaches to processing online student activity data group student actions into smaller sessions. %
In the case of click actions, these sequences are generally referred to as clickstreams.
The goal when breaking a flow of actions into session clickstreams is to maintain some notion of interpretability, i.e. to devise meaningful study sessions.
While this appears to be easy in some cases, it is generally non-trivial to find automated cut-offs rules that find sensible representations of study sessions for a large and diverse set of clickstreams at once.

Previous research suggests several different strategies to split clickstreams.
The authors of \cite{brooks_time_2015} choose fixed duration time frames to group student actions from a several months long MOOC. The researchers decide for durations
between one day and one month and show some success in the downstream prediction of student achievements with their choices.
Similar fixed durations are used in \cite{amnueypornsakul_predicting_2014}.
Another popular splitting strategy is based on time-out thresholds where a new sub-session is started when no action was performed in a predefined time window \cite{marquardt_pre-processing_2004,ba-omar_framework_2007,munk_impact_2011,del_valle_online_2009,desmarais_clustering_2013}.
The authors of other studies go one step further and combine the approaches by first, splitting at a fixed duration cut-off and second, at data-driven timeout thresholds of 15 minutes for `study sessions' and 40 minutes for `browser sessions' \cite{gitinabard_what_2019}.
Similar data-driven approaches are pursued in \cite{wise_broadening_2013, sheshadri_predicting_2018}.
Other common heuristics include splitting at navigational criteria such as reloading of the course page \cite{kovanovic_penetrating_2015}.

On a high level, the problem of devising meaningful sub-sessions is closely related to the problem of time-at-task estimation in web-usage mining. %
Ideally, study sessions reflect time periods in which students interact with the material without any major breaks or distractions.
There is a rich body of literature on time at task estimation that suggests that there is no one-fits-all solution to finding suitable time windows to split activity streams at \cite{kovanovic_penetrating_2015, baker_press_2007,cooley_data_1999}.
Previous research suggests that the exact splitting heuristic can have a significant effect on overall model fit, model significance, and even interpretation of findings in the downstream modeling tasks \cite{kovanovic_penetrating_2015}.
In \cite{kovanovic_penetrating_2015}, the authors explore the effect of 15 different time-at-task estimation procedures on five different models of student performance.
Overall, the authors conclude that there is no universally best method and recommend a mixture of existing methods including data-driven components.
Following this suggestion, we employ a multi-step splitting procedure including navigational criteria, data-driven time-out thresholds, and separation of assessment weeks inspired by the procedure in \cite{gitinabard_what_2019}.

\subsection{Sequence Analysis}

Different methods have been proposed to process sequence-type student action data dependent on the amount of data, the length of sequences, and the goal at hand.
Several lines of research rely on Markov chains and hidden Markov models which lend themselves well to visualization of sequences, but can make quantification of group differences in outcomes challenging \cite{geigle_modeling_2017,faucon_semi-markov_2016,jeong_mining_2008}.
Another commonly used class of methods is clustering of activity sequences \cite{desmarais_clustering_2013,kizilcec_deconstructing_2013,guerra_problem_nodate}.
Using data from three large MOOCs, the authors of \cite{kizilcec_deconstructing_2013} draw on simple $k$-means clustering of sequences of interactions with video lectures and assessments and observe four high-level student trajectories: completing assessments, auditing the course, disengaging after a while, and sampling content.
In order to cluster the sequences, the authors rely on a numerical translation of student actions. 
The authors of \cite{desmarais_clustering_2013} cluster and visualize students' interactions with a college math environment, and instead rely on Levenshtein distance to measure the distance between sequences. %
Some works combine Markov models and clustering to account for the randomness introduced by the Markov models and report more robust results \cite{shih_unsupervised_2010,kock_activity_2011,klingler_temporally_nodate}.

Although the described methods allow for a relatively easy grouping of sequences, interpretation of clusters can be non-obvious. One way to address this problem is to deliberately focus on finding relevant sub-parts of action sequences. Methods based on this goal can be summarized under the term pattern mining, and are both wide-spread and diverse.
A relatively recent approach is given by differential pattern mining which focuses on automatically extracting patterns that are both above a certain threshold in frequency, and sufficiently different among groups of interest (e.g. high and low achieving students) \cite{kinnebrew_contextualized_2013, kinnebrew_identifying_nodate}. %
Other lines of research rely on more traditional data mining techniques \cite{herold_mining_nodate, mukala_exploring_2015}, or extraction of $n$-grams, i.e. sub-sequences of $n$ consecutive actions \cite{brooks_time_2015,martinez_analysing_nodate,wen_identifying_2014,pardos_analysis_2017}.
The authors of \cite{martinez_analysing_nodate} use a multi-step procedure to extract frequent $n$-grams that are subsequently used to identify different strategies in a collaborative interactive tabletop game. 
Part of our analysis is based on a similar approach to extract frequent behavioral patterns, and combines ideas of $n$-gram extraction and clustering to get more robust results.

A different class of promising methods is rooted in Natural Language Processing (NLP).
Hybrid language models lend themselves well to the sequential structure of education data, and their use for student activity sequences has lead to some success in retrieving patterns and creating new visualizations. The underlying idea is that, given sufficiently fine-grained data, students' sequential actions resemble words building sentences and can be attributed some `semantic meaning'.
The NLP toolbox has not yet been explored fully, but some attempts to using language models for educational data are noteworthy and relevant for the context of our work.
The authors of \cite{wen_identifying_2014} use topical $n$-gram models to automatically extract `topics' in the form of frequent patterns from clickstreams.
In \cite{pardos_analysis_2017}, the authors train a skip-gram neural network to receive a structure preserving vector embedding of the types of clicks student can make.
After standard dimensionality reduction, the researchers are able to provide a new kind of visualization of students' trajectories through the course. %
Since modern NLP models generally require large amounts of granular training data, work relying on these models has exclusively focused on MOOCs so far.
In this study, we draw on Latent Dirichlet Allocation (LDA) in order to automatically extract frequent patterns and compare derived student strategies against the results of a more traditional $n$-gram pipeline.
In some sense, LDA is similar to the ideas proposed by \cite{wen_identifying_2014} but requires less training data which renders it particularly useful for blended courses.
In addition, we use an adapted form of the skip-gram model proposed by \cite{pardos_analysis_2017} in order to explore the context of student actions in our data.
To the best of our knowledge, this is one of the first works to employ NLP methods for analysis of blended courses.

\section{Data}
\label{sec:data}

\subsection{Data Context: Diderot}
\label{sec:diderot}

The data this study builds on was collected through the educational software system Diderot.
Diderot is a cloud-based course support system commonly used to assist undergraduate and graduate level college courses.
The system spans a wide range of functionalities including sharing of lecture notes, a discussion board (called post office), in-class attendance polls, homework submission, and automatic code grading.
This bandwidth usually renders the use of additional outside technological course support unnecessary. In turn, the student usage data collected from Diderot can give an almost comprehensive view on students' course participation outside of face-to-face class times.

\begin{figure}
	\centering
	\includegraphics[scale = 0.53]{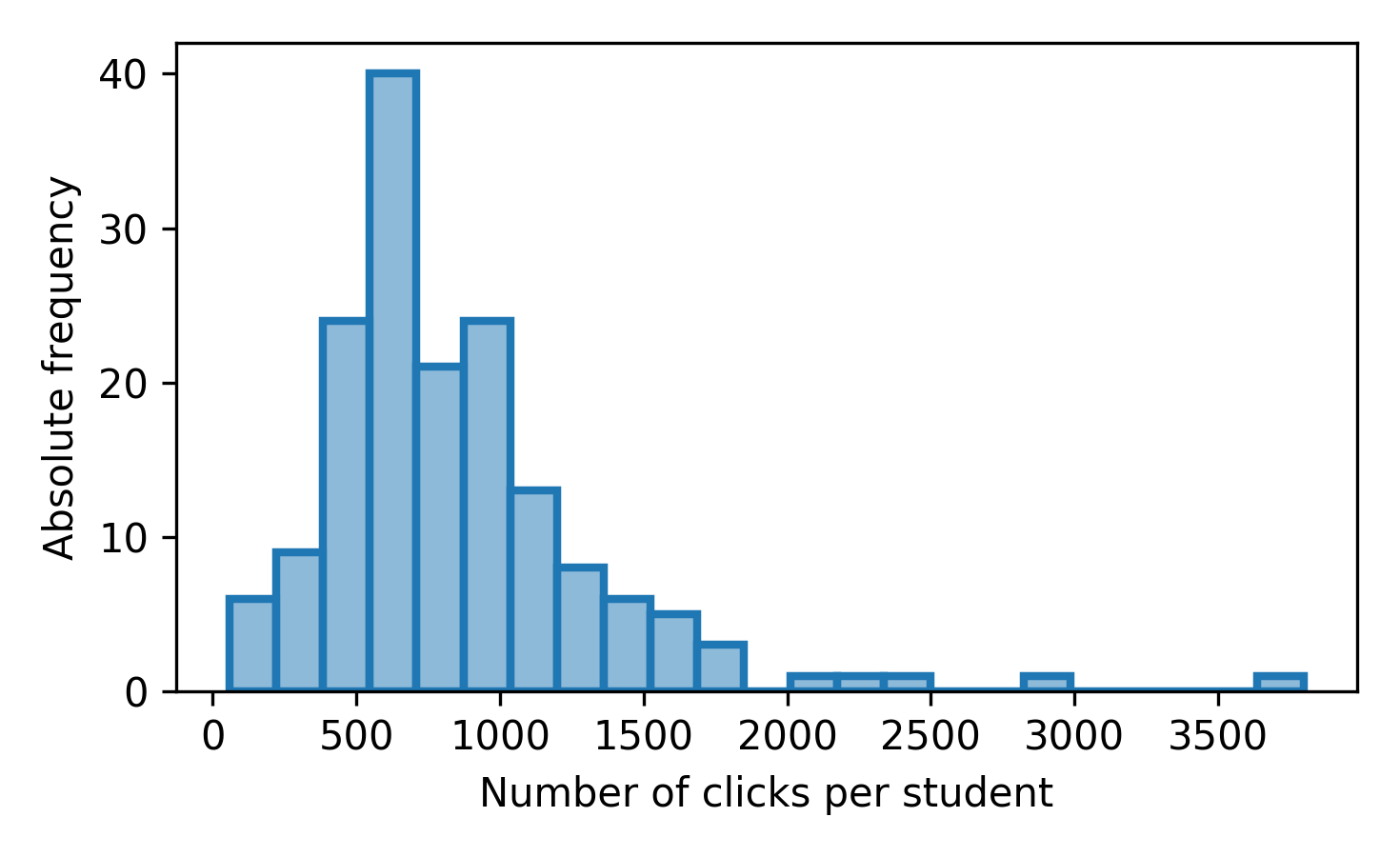}
	\includegraphics[scale = 0.53]{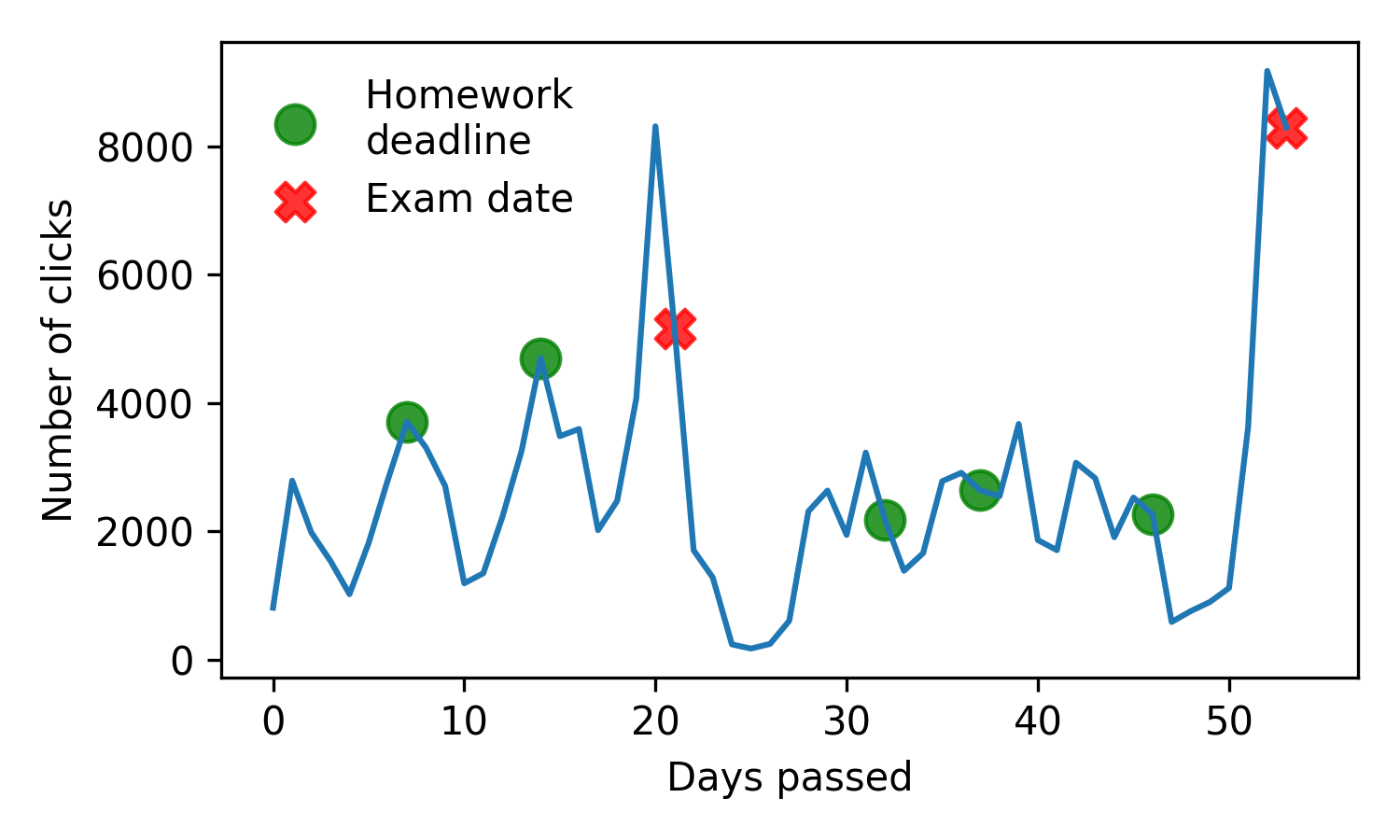}
	\includegraphics[scale = 0.53]{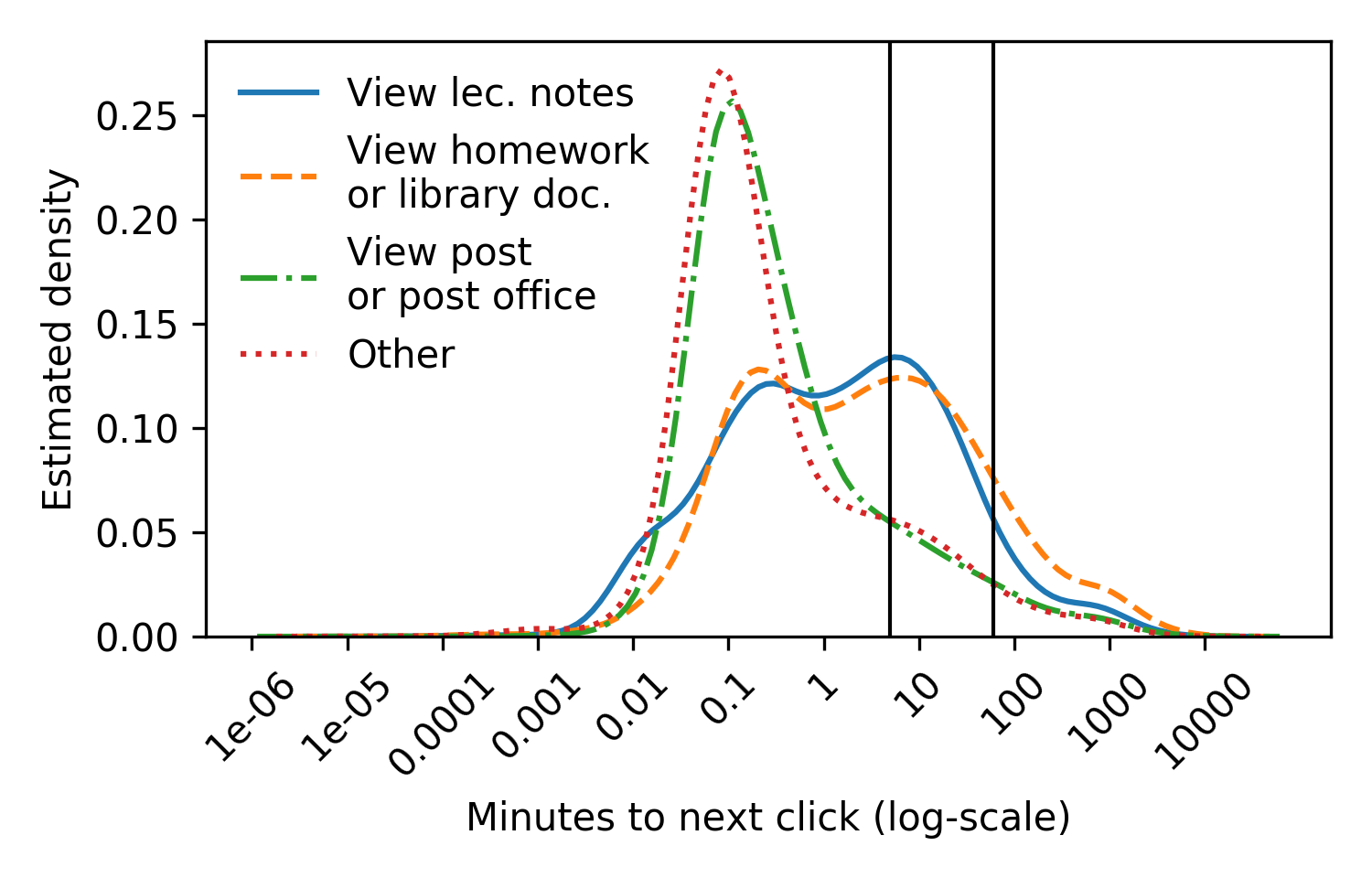}
	\caption{Histogram of the number of clicks per student. We observe 138,960 clicks spread between 164 students (top).
		Number of clicks over observation period with assessment deadlines highlighted (middle).
		Kernel density estimates for log-distribution of waiting times between clicks dependent on type of last click after splitting at assessment weeks and \texttt{Load course} actions. Final cut-offs at 5 and 60 minutes are indicated by vertical lines (bottom).}
	\label{fig:data}
\end{figure}

When it comes to sharing of lecture notes, Diderot takes a more granular and interactive approach as compared to traditional learning management systems.
Content is split into small sub-entities (called atoms) which are displayed in a linear fashion following the outline of a chapter.
Atoms are highly interactive and come with a variety of clickable icons that allow students to take notes, bookmark, follow, or like atoms, and, in particular, to ask questions concerning their content.
Discussions about course material that are sparked in this way are visually attached to the respective atom, allowing other students to submit comments.
This setup results in much richer data on interactions with lecture notes than we can expect from PDF formatted lecture material.

Most student usage data from Diderot is presented by individual, time-stamped click actions that come with various identifiers coding the exact type, user, and location of the interaction. %
In turn, the activity data can be broadly separated into navigation (e.g. \texttt{Load course}, \texttt{Click link}, \texttt{Search}), discussion (e.g. \texttt{Go to post office}, \texttt{Create post}), and behaviors (e.g. \texttt{Like atom}, \texttt{Follow post}).

\subsection{Data Description And Exploration}
\label{sec:datadescription}

Our data is drawn from the second half of a large sophomore-level computer science course taught at Carnegie Mellon University in spring 2019.
Since data is not available for the first part of the course due to initial technical difficulties, we exclude all students who dropped the course throughout the semester.
One additional student was excluded based on inflated click patterns which suggested an attempt at automatically scraping content.
Along with the click data, we rely on performance information measured by homework and exam grades, as well as student-level lecture and recitation attendance logs.
All data is collected through Diderot and matched based on anonymous student identifiers.
A summary of the click data over the seven week observation period is displayed in Figure~\ref{fig:data}.

\textbf{Types of clicks.} 
At finest granularity, Diderot allows for several tens of thousands distinct click actions within a single course since every individual click is associated to a fully specified object and activity.
However for the sake of analysis, we group clicks into different types where the appropriate level of granularity is non-obvious.
We aggregate clicks based on the type of object they refer to as well as the activity performed.
In order to maintain interpretability, this aggregation is performed separately in each sub-part of the course given by lecture notes, homework material, recitation notes, a library documentation (which is comprised of coding references), and practice exams.
This leaves us with 37 different click types, the most common of which are summarized in \Cref{table:screenids}.

\textbf{Grades and types of assessment weeks.}
Performance outcomes are measured by percentage grades in five homeworks and two exams (a midterm and final exam) that fall into the observation period.
This naturally divides the data into seven assessment weeks with a deadline for a homework problem set or exam at the end of each period. 
Deadlines are approximately evenly spaced with only one extended homework period of 11 days after the midterm exam (which also spans over a four day spring holiday), followed by a shorter homework period of only 5 days.
We take interest in relating students' study behavior to two distinct outcome variables: (1) The type of the assessment week, i.e. homework deadline or exam, and (2) the percentage grade students received in the respective assessment.
As depicted in \Cref{fig:data}, there are visible spikes of increased activity before the assessment week deadlines especially before the two exams.
In addition, we note that the distribution of grades appears notably different between homeworks and exams which is confirmed by a two-sample Kolmogorov-Smirnov test ($p<0.001$).
While the distribution of exam grades is approximately bell-shaped with heavy tails and a slight left-skew, i.e. more particularly high scores than particularly low scores, the homework grade distribution is left-skewed with additional modes at 0 and 100.
This difference in distributions is unsurprising as exams are generally graded on a curve and cannot be skipped by students, while homeworks allow for more variability.

\begin{table}
	\centering
	\caption{Summary of the most frequent click types.} %
	\label{table:screenids}
	\small
	\setlength{\extrarowheight}{3px}
	\begin{tabular}{|lrr|}
		\hline
		\textbf{Click type} & \textbf{Count}&\textbf{Share}\\
		\hline
		View chapter in lecture notes & 24,420 & 17.57\,\%\\
		View general post & 21,555 & 15.51\,\%\\
		Load course & 19,677 & 14.16\,\%\\
		View post office & 16,231 & 11.68\,\%\\
		View atom post& 15,468 & 11.13\,\%\\
		View homework atom & 7,888 & 5.68\,\%\\
		\hline
	\end{tabular}
\end{table}

\textbf{Class attendance.}
Attendance in lecture and recitation sessions was taken with Diderot polls. If a student participated in the poll, which was generally only open for a few minutes, it was assumed that they attended the session. We treat attendance in lectures and recitations separately and aggregate the binary information on an assessment week basis by taking the mean. In turn, student's attendance scores lie between 0 and 1 with the exception of the final exam week which is not associated to any contact class time.

\section{Methods}
\label{sec:methods}

\subsection{Session Clickstreams}

In raw form, each student is associated with a single clickstream which consists of ordered click actions over the whole observed time period.
We employ a multi-step procedure to split this data into more meaningful study sessions.
First, we divide the clickstreams based on assessment weeks.
Second, we split the resulting sub-clickstreams each time a \texttt{Load course} action is recorded, and last, 
we choose a data-driven timeout threshold to further break up the resulting sequences.

In order to find a suitable timeout threshold, we employ a technique similar to \cite{gitinabard_what_2019} and examine the distribution of time differences in the sub-sequences.
We find that the distribution of waiting times supports a wide range but is rapidly decaying.
While 75\,\% of clicks are made within 2.81 minutes or less, a small subset of clicks has time differences of up to 7 days.
\Cref{fig:data} shows kernel density estimates of log-transformed minutes until the next click within the sub-clickstreams obtained after the second step of our procedure.
Different estimates are obtained for distinct categories of actions.
While the logarithmic distribution of post-related and miscellaneous clicks is unimodal with the majority of follow-up clicks made within one minute, the distribution for clicks related to homework and lecture notes has an additional mode at about 5-10 minutes.
This disparity is unsurprising given that most actions can be expected to be short, while reading through lecture notes or homeworks can be a more lengthy process. 
In order to preserve both types of sessions, we separate clickstreams at a 60 minutes threshold if the last action was loading of lecture notes or homework related content, and at 5 minutes otherwise.
As a result, we obtain a total of total of 35,703 session clickstreams where each clickstream has between one and 115 clicks with mean of 3.98 clicks and standard deviation of 6.23; 75\% of session clickstreams have at most 4 clicks.

\subsection{Context Of Click Types}
\label{sec:context}

\begin{figure}
	\centering
	\includegraphics[scale=0.36]{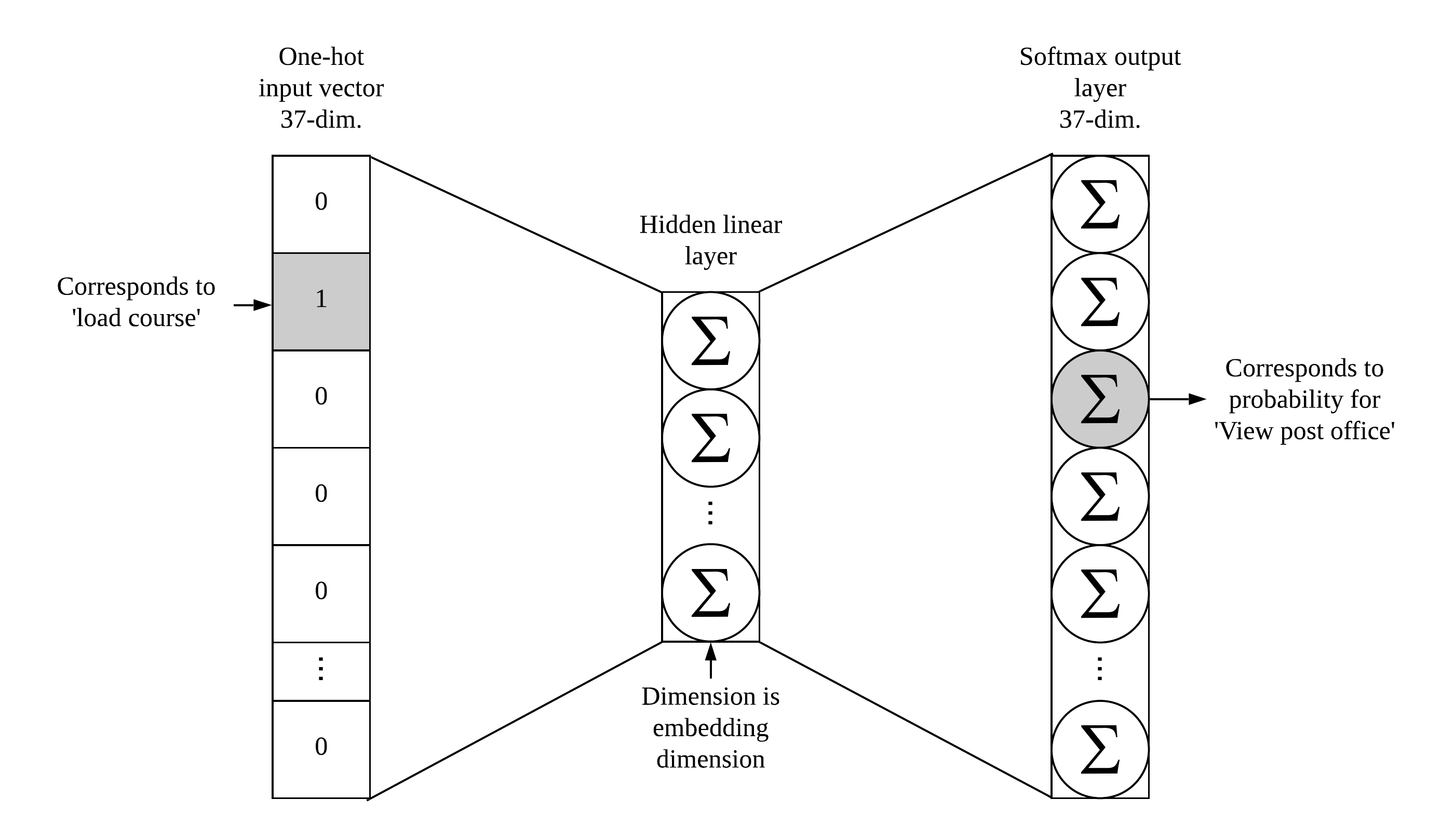}
	\caption{Skip-gram neural network. The hidden layer linearly transforms one-hot encoded inputs while the softmax output layer approximates the probability that each given click type appears in the same context as the input click. After training, the weights of the hidden layer provide a structure preserving embedding of click types.}
	\label{fig:skipgram}
\end{figure}

We explore the contexts in which different types of clicks are made in order to gain some understanding of how students generally use the course support system.
This is crucial since Diderot is a fully integrated interactive platform that allows the same type of click in contexts that can have different interpretations.
Inspired by \cite{pardos_analysis_2017}, 
we tackle this problem by devising a structure-preserving embedding of the click types into a real-valued vector space, i.e. each click type is mapped onto a vector such that click types that appear in the same contexts or are interchangeable are close to each other. 
This type of embedding can be obtained from a skip-gram model which is a common supervised two layer neural network model often used for language type data (see Figure~\ref{fig:skipgram}).

Training data for the model is build by extracting pairs of neighboring click types from the session clickstreams. More concretely, each input click is paired with each click appearing within some index in the same clickstream.
Both the window size and the number of hidden units are important hyperparameters. 
Since most of our clicks are short and we seek an embedding of only 37 clicktypes, we explore small values for both parameters, i.e. window sizes in $\{1,2\}$ and embedding sizes in $\{3,4\}$. After this small grid search, we only retain the model with the lowest average training loss in the last 2000 training steps. 
In order to speed up training, we rely on mean noise-contrastive estimation (NCE) loss where 8 negative classes are sampled for every batch instead of computing the entire softmax output.
All models are trained over a maximum of 300,000 training steps with SGD with learning rate 1 and a batch size of 512. Training is terminated early when the average loss over 2000 training steps does not change considerably for 5 consecutive non-overlapping 2000-step periods.
Because training the model is only the surrogate task in order to obtain the embedding, we train on all available data which comprises 206,514 or 363,260 pairs dependent on the window size.

\subsection{Frequent Pattern Extraction}
\label{sec:freqpatterns}

\subsubsection{Clustered $n$-grams}

We refer to finite sub-sequences of clickstreams as frequent patterns if they appear various times across different students, study sessions, and assessment weeks.
Our goal is to automatically extract frequent patterns which represent some kind of strategy or high level task students are fulfilling.
As an example, the sequence \texttt{[Login -- View post office -- View general post]} could be interpreted as an attempt to catch up on the course news. 
Pattern mining in educational data mining can lead to relatively unstable results. 
In order to increase robustness, we examine and compare the results of two distinct procedures for frequent pattern extraction.
The first method resembles the procedure proposed by \cite{martinez_analysing_nodate}, and consists of a multi-step procedure which first extracts a large set of candidate patterns, and then narrows the selection down by similarity grouping.
Formally, we proceed according to the following steps:

\vspace{-2ex}
\begin{enumerate}
	\item [(1)] \textbf{All $n$-grams.} We extract $n$-grams, i.e. consecutive sub-sequences of $n$ clicks, from the session clickstreams.
	 Since we expect very short patterns to be uninterpretable, and particularly long patterns are rare in our dataset, we choose $n=3,4,5$.
	\item [(2)] \textbf{Candidate patterns.} Only the most frequent patterns are kept as candidates for further analysis. Following some experimentation, we choose to keep the most frequent 1\,\% of patterns of each length.
	\item [(3)] \textbf{Hierachical clustering.} The set of candidate patterns can be expected to be repetitive in the sense that patterns might be similar but vary in length or differ in a single click action but yield the same interpretation.
	To address this issue, we automatically group candidate patterns by agglomerative clustering with average linkage. The number of clusters, and thus of final frequent pattern categories, is chosen by visual inspection of the model's dendrogram. 
\end{enumerate}
\vspace{-2ex}

The final step of this procedure requires us to specify a notion of similarity between patterns.
In some sense, it is natural to draw on a string distance measure as sequences of clicks resemble many of the characteristics we would expect from natural language.
While the authors of \cite{martinez_analysing_nodate} draw on the traditional Levenshtein distance, 
we choose the Jaro-Winkler distance between two patterns $p_1,p_2$ measured by $1-jw(p_1,p_2)$, where $jw(\cdot,\cdot)$ denotes the Jaro-Winkler similarity. Jaro-Winkler distance is an adaptation of more traditional edit distances which takes the sequence length as well as common starting sub-sequences into account.
This allows more sensible measuring of similarities between repetitive patterns of different lengths such as the 3-gram \texttt{[View general post -- View general post -- View general post]} and the 5-gram \texttt{[View general post -- View general post -- View general post -- View general post -- View general post]}. Intuitively, the two patterns should have a low distance and in fact, their Jaro-Winkler distance is approximately 0.093 while their normalized Levenshtein distance is 0.4. %
For our purpose, we treat each click as a character that can be exchanged or transposed for a penalty on the distance. 
Then, the Jaro similarity is defined as
$$
	j(p_1,p_2):=
	\begin{cases}
	0 \text{ if } m = 0, \\
	\frac{1}{3}\left(\frac{m}{\lvert p_1\rvert}+\frac{m}{\lvert p_2\rvert}+\frac{m-t}{m}\right) \text{ else,}
	\end{cases}
$$
where $m$ is the number of matching clicks within an index window of $\lfloor (\max\{\lvert p_1 \rvert, \lvert p_2 \rvert\}/2) \rfloor -1$, and $t$ is half the number of required transpositions for matching clicks.
Further, the Jaro-Winkler similarity is defined as 
$$
	jw(p_1,p_2) := j(p_1,p_2) + \frac{l}{10}(1-j(p_1,p_2)),
$$
where $l$ is the length of a common starting sequence between $p_1$ and $p_2$ (at most 4).
The additional scaling ensures that distances are normalized to lie in $[0,1]$.

\subsubsection{Topic Model}

The clustered $n$-grams procedure of extracting frequent patterns is easy to implement and model-free.
However, it requires us to choose several hyperparameters such as the size of $n$-grams, the share of candidate patterns, or the number of clusters.
It is also likely that the exact choice of the edit distance in the clustering step has a non-negligible effect on the observed results.
In order to test our results for robustness, we employ a second method for pattern extraction and compare the resulting student strategies.
This method draws on the idea that session clickstreams resemble sentences, individual clicks resemble words, and there is some notion of semantic to a sequence of clicks.
Based on these similarities, we use Latent Dirichlet Allocation (LDA), a common NLP model that allows automatic extraction of topics from written documents.
LDA is a Bayesian model which, in our case, is build on the assumption that each session clickstream is a mixture of patterns and each pattern is a mixture of clicktypes.
We use the words pattern and topic interchangeably here. 
While the clickstreams (and hence clicktypes) are given to the model, the topics are latent and can be inferred from the fitted model.
The prior on the session clickstream generation assumes that $M$ clickstreams of lengths $N_1,\ldots,N_M$ are drawn according to the following steps.
(1) Draw a topic distribution $\theta_i\sim Dir_k(\alpha)$ for each $i=1,\ldots,M$, where $k$ is the number of topics.
(2) Draw a click type distribution for topics $\phi_i\sim Dir_V(\beta)$ for each $i=1,\ldots,V$, where $V$ is the number of different click types.
(3) For each click position $i,j$ with $i\in\{1,\ldots,M\}$ and $j\in\{1,\ldots,N_i\}$, first, choose a topic according to $z_{ij}\sim Multinomial(\theta_i)$, and second, draw a click type from $w_{ij}\sim Multinomial(\phi_{z_{ij}})$.
LDA comes with three hyperparameters: the prior Dirichlet parameters $\alpha$ and $\beta$ which express some prior belief on how the mixtures of topics and click types are composed, and the number of latent topics $k$.
While we set the prior Dirichlet parameters to suggested default values, i.e. normalized asymmetric priors, the number of latent topics requires some more thought.
Recent research suggests the use of topic coherence measures for comparison of models with different choices of $k$ \cite{mimno_optimizing_nodate,stevens_exploring_2012}.
On a high level, topic coherence attempts to measure semantic similarity between high scoring words (or here click types) in each topic which gives some indication of how interpretable the topics in question are.
We experiment with several numbers of topics ranging around the number of frequent patterns extracted by the clustered $n$-gram technique. Since no significant differences in coherence can be observed, we resort to using the same number of topics as for the clustered $n$-gram method for the sake of comparison.

\subsection{Prediction Models}

\textbf{Frequent patterns counts as features.}
In order to explore what role the extracted strategies play in homework solving versus exam preparation and whether they drive success, we build two prediction models based on patterns counts from the clustered $n$-gram method.
For this, a representative pattern of 3 clicks is chosen for each of the devised strategy clusters, and its occurrences in each of the session clickstreams is counted by comparing against each 3-gram derived from the clickstream.
Since we cannot expect the chosen pattern to accurately represent the whole cluster, we allow a Jaro-Winkler distance up to 0.2 when comparing the sub-sequences.
This procedure allows matching of click sequences with only one replacement ($1-jw(abc, abd) \approx 0.18$), one transposition ($1-jw(abc, acb) \approx 0.10$), or one replacement and one trasposition ($1-jw(abc, adb) \approx 0.20$).
In order to build student and assessment week based prediction models, we aggregate pattern counts along assessment weeks and individual students by simple addition. 
Similar methods have been employed by \cite{brooks_time_2015,li_2016,sinha_2014,coleman_2015}.

\textbf{Predicting assessment type.}
A random forest classifier is trained to predict the assessment type, i.e. homework or exam, from frequent pattern counts, the number of clicks, and the number of session clickstreams a student has within a given week.
In practice, it is unlikely that we would need to predict the assessment type as it is usually known. However when paired with careful analysis of feature importance and partial dependence, such model can yield valuable insights into the most important differences in student behavior between homework and exam weeks. 
We use 80\,\% of the 1,148 student-week combinations for training and hold back 20\,\% as test set.
Hyperparameters including the maximum tree depth, the maximum number of features to consider at splits, the minimum number of samples per leaf, and the number of trees are chosen by a grid search over a range of values, where models are trained with 5-fold cross validation on the training set. 
Our model draws on Gini impurity to measure the quality of splits, and we evaluate feature importance based on the mean decrease in impurity (MDI) associated with splitting at a given feature when predicting $Y$.
For a set of fitted trees $\mathcal{T} = \{T_1,\ldots,T_N\}$, the MDI of a feature $X_m$ is defined as
\begin{align}
	\label{eq:mdi}
	MDI(X_m) = \frac{1}{N}\sum_{T\in\mathcal{T}} \sum_{t\in T: v(s_t) = X_m} p(t)\Delta i(s_t,t),
\end{align}
where $p(t)$ is the proportion of samples that reaches node $t$, $v(s_t)$ is the variable used to split $s_t$, and $\Delta i(s_t,t)$ is the decrease of impurity generated by the split.

\textbf{Predicting grade outcomes.}
Similar to the assessment type prediction model, we train a random forest regressor to predict students' grade outcomes based on strategy counts, the number of clicks, the number of session clickstreams, and attendance information.
The additional consideration of lecture and recitation attendance requires us to remove all observations from finals week, since no face-to-face class time has taken place in the last week of the course. Since this constitutes half of all exam observations in the data and grade distributions of homeworks and exams are significantly different ($p < 0.001$), we limit our prediction model to homework grade prediction entirely.
Of the 820 homework samples, 80\,\% are used for training and 20\,\% for testing. 
A grid search of hyperparamters with 5-fold cross-validation on the training set is performed, and
feature importance is measured analogous to Equation~\ref{eq:mdi} with the MSE as impurity measure.

\section{Results}
\label{sec:results}

\subsection{RQ1 How do students interact with course material, and what are frequent strategies they take?}

\begin{figure}
	\centering
	\includegraphics[scale = .45]{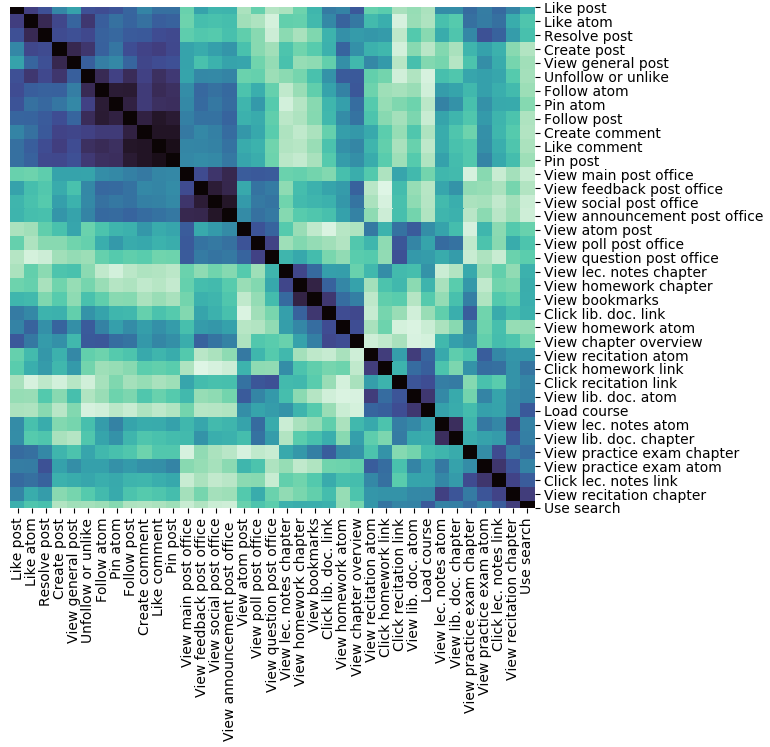}
	\caption{Euclidean distances of click type embeddings based on skip-gram neural network.
		Darker color suggests that embedding are close. Proximity in the embedding space suggests clicks generally appear in similar contexts. Rows and columns are clustered for visualization.}
	\label{fig:skip_gram}
\end{figure}

\subsubsection{Context Of Click Types}

In order to gain some initial understanding of online student behavior, we explore the contexts in which different types of actions are performed by deriving a skip-gram neural network based embedding of actions.
After exploring a small grid of hyperparameter values, our skip-gram is trained on data pairs with window size 1 to learn a 4-dimensional embedding.
\Cref{fig:skip_gram} depicts the Euclidean distances between the embedding vectors of different click types based on the model.
Proximity of embeddings suggests that click types either appear in a similar context, i.e. within a few clicks of each other, or are interchangeable actions, i.e. have the same context.
In other words, by exploring which actions lie close to a given click type in the embedding space, we can gain some insight into the set of clicks students typically make right before and after. %
It is noteworthy that some types of actions appear together by design of the Diderot system, e.g. in order to comment on a post, the post has to be loaded. \Cref{fig:skip_gram} reflects many of these expected relations which gives some validation to our methodological approach.

Our results suggest several broad clusters of student actions.
The block in the upper left corner of \Cref{fig:skip_gram} appears to focus on active discussion participation including click types such as \texttt{Like post} or \texttt{Create comment}.
The next block is somewhat close to many of the active discussion actions and concentrates on scrolling through the discussion board represented by \texttt{View post office} type actions.
Although more rigorous statistical analysis is needed, the results suggest some interesting interpretations:

\vspace{-2ex}
\begin{enumerate}
\item [(1)] \textbf{Students ask more questions about homeworks than about any other course materials.} This interpretation is based on the proximity of \texttt{Create post} to \texttt{View homework atom} which appears to be much closer than any other \texttt{View atom} type action.
This suggests that student questions, comments and clarifications are more common for homework material than for lectures notes, recitation material, practice exams, or the library documentation.
\item [(2)] \textbf{Students are more likely to interact with course-wide posts than material specific discussions.}
The action \texttt{View general post} is close to interactive behavior such as \texttt{Create comment}, \texttt{Like post} or \texttt{Follow post} while \texttt{View atom post} appears to be performed mostly in a different context.
This suggests that discussion-specific reactions and interactions concentrate mostly on general posts such as course announcements or social posts and are less common for questions and comments concerning particular parts of the course materials.
\end{enumerate}
\vspace{-2ex}

Overall, context analysis for click types based on skip-gram neural networks provides us with some valuable understanding of students' use of Diderot.
The same method might be useful to other practitioners, in particular, for initial exploration of data collected through educational software systems. 
It appears that interpretable low-dimensional embeddings of a medium number of action types can be obtained with only a few weeks worth of data from a a single college course which renders this method particularly useful for blended courses.

\subsubsection{Frequent Pattern Extraction}

\begin{table*}
	\centering
	\caption{Comparison of student strategies extracted by clustered $n$-gram method and LDA. Patterns in the first block (B1) consist of exactly the same click types, while other patterns show differences but allow for similar interpretations (B2). Lastly, the LDA method finds a mixture of practice exam related patterns and a new load course pattern (B3).}
	\label{table:topics}
	\small
	\setlength{\extrarowheight}{3px}
	\begin{tabular}{|L{.3cm}|L{2.85cm}L{4.9cm}|L{2.85cm}L{4.9cm}|}
		\hline
		&\multicolumn{2}{c|}{\textbf{Clustered $n$-gram method}} & \multicolumn{2}{c|}{\textbf{LDA method}} \\
		\hline
		&\textbf{Student strategy} & \textbf{Associated click types} & \textbf{Student strategy} & \textbf{High weight click types}\\
		\hline 
		B1&Look at lecture notes & View lecture notes chapter (75.88\,\%) & Look at lecture notes & View lecture notes chapter (1)\\
		&Look at homeworks & View homework chapter (100\,\%) & Look at homeworks & View homework chapter (0.826)\\
		&Look at recitation material & View recitation chapter (100\,\%) & Look at recitation material & View recitation chapter (0.712)\\
		\hline
		B2&Catch up on news & View general post (52.04\,\%), View main post office (24.3\,\%), View atom post (16.13\,\%) & Catch up on news & View general post (0.543), View main post office (0.410)\\
		&Active homework engagement & View atom post (50\,\%), View homework atom (31.02\,\%), View general post (10.65\,\%) & Active homework engagement & View atom post (0.653), View homework atom (0.344)\\
		&In-depth review of lecture notes & View lecture notes atom (50\,\%), View atom post (28.57\,\%), View lecture notes chapter (21.43\,\%)& In-depth review of lecture notes & View lecture notes atom (0.483), View atom post (0.31), Click link lecture notes (0.195)\\
		&Look at library documentation & View library documentation chapter (85.29\,\%) & Look at library documentation & View library documentation chapter (0.674), Search atom (0.321)\\
		\hline
		B3&Go through a practice exam & View practice exam atom (100\,\%) & Practice exams & View practice exams chapter (0.658), View practice exams atom (0.341)\\
		&Look at practice exams & View practice exams chapter (100\,\%) & Load course & Load course (0.998)\\
		\hline
	\end{tabular}
\end{table*}

Patterns are extracted with two distinct methods, and subsequently interpreted in terms of underlying student strategies. A summary of the results and comparison between the methods is given in \Cref{table:topics}.
The left side of the table shows the results of the clustered $n$-gram pipeline for pattern extraction. The most frequent 1\% $n$-grams for each $n=3,4,5$ are extracted from the session clickstreams. This yields a candidate set of 223 sequential patterns which are clustered into 9 groups based on agglomerative clustering with average linkage and Jaro-Winkler distance as distance function. The number of clusters is informed by visual inspection of the respective dendrogram.
It is noteworthy that the clusters appear to have imbalanced sizes with the largest cluster including 106 candidate patterns, and the smallest clusters containing only 2 or 3 of the candidate patterns. Yet, inspection of the associated click types and their in-cluster frequencies allows for intuitive interpretations as student strategies.
Multiple of the devised strategies revolve around passive review of materials such as lecture notes, homeworks, recitation material, library documentation (which includes code snippets for reference), or practice exams.
More involved strategies are given by active homework engagement, in-depth review of lecture notes, catching up on course news, and going through practice exams. For example, the catching up on course news strategy is associated with sequential patterns involving reading of general posts, atom posts, and loading the main post office page.

The right side of \Cref{table:topics} summarizes the results of pattern extraction based on Latent Dirichlet Allocation (LDA). For the sake of comparison, we keep the number of extracted patterns fixed and derive 9 student strategies.
By assumption of the model, each pattern is a mixture of all click types. In turn, extraction of weights is straightforward and we report the click types with highest weights for each pattern.
We find that multiple of the extracted patterns match exactly the patterns retrieved with the clustered $n$-gram method in the sense that they are based on exactly the same click types (B1).
Another set of patterns shows small changes in included click types, but essentially provides the same interpretation as the patterns found with the first method (B2).
Lastly, the LDA method finds a practice exam strategy which broadly presents a mixture of the two practice exam related strategies from the first model, and a load course strategy which almost entirely consists of the \texttt{Load course} action (B3).
The load course pattern likely arises from the session clickstreams with a single click which present 30.30\% of the session clickstreams. A total of 56.66\% of these one-click sequences are \texttt{Load course} actions.
Reasons for these single \texttt{Load course} clicks can be manifold. In some cases, students might get distracted immediately after loading the course, or they have to reload the course multiple times. However, we hypothesize that in most cases, the course overview page which is loaded when loading the course provided all information the student was looking for since it includes recent updates, posts and announcements.
Contrary to the clustered $n$-gram method which only takes into consideration session clickstreams of at least three clicks, LDA can leverage even these short clickstreams. %
Yet, the additional insights gained through the load course pattern are marginal since it very short and hard to interpret as a strategy.

All in all, both methods roughly extract the same strategies which speaks in favor of the validity of both approaches.
One could argue that the clustered $n$-gram method yields slightly more tangible insights since the patterns present actually frequently occurring sub-sequences. 
However for larger data sets, the method can become computational expensive rendering LDA a better choice.

\subsection{RQ2 How do students use these strategies for homework solving as compared to exam preparation?}

\begin{figure}[th]
	\centering
	\includegraphics[scale=.52]{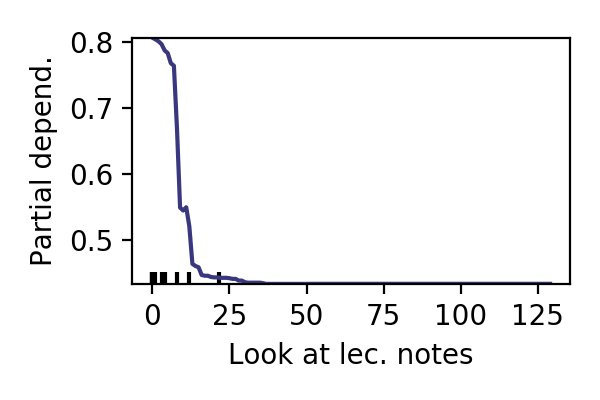} 
	\includegraphics[scale=.52]{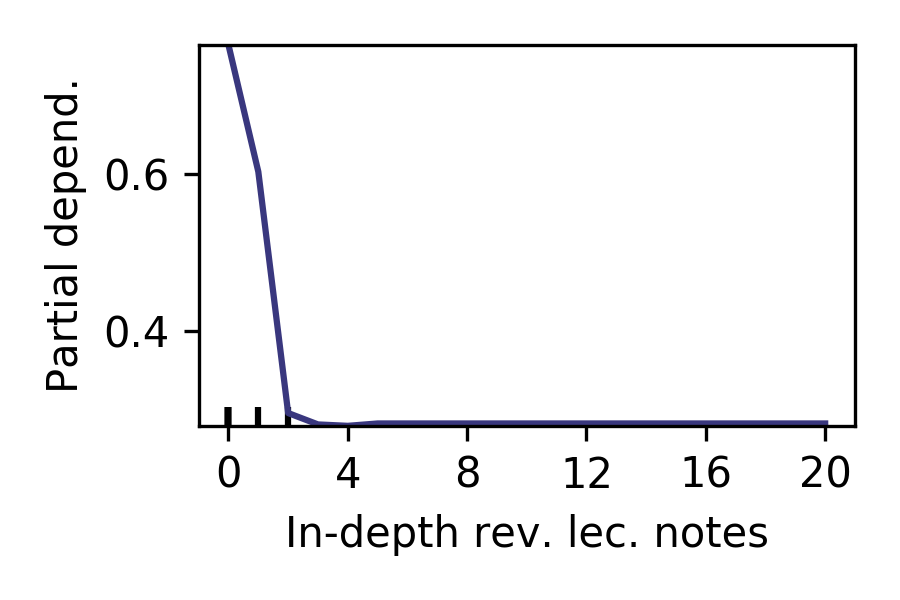} \\
	\includegraphics[scale=.52]{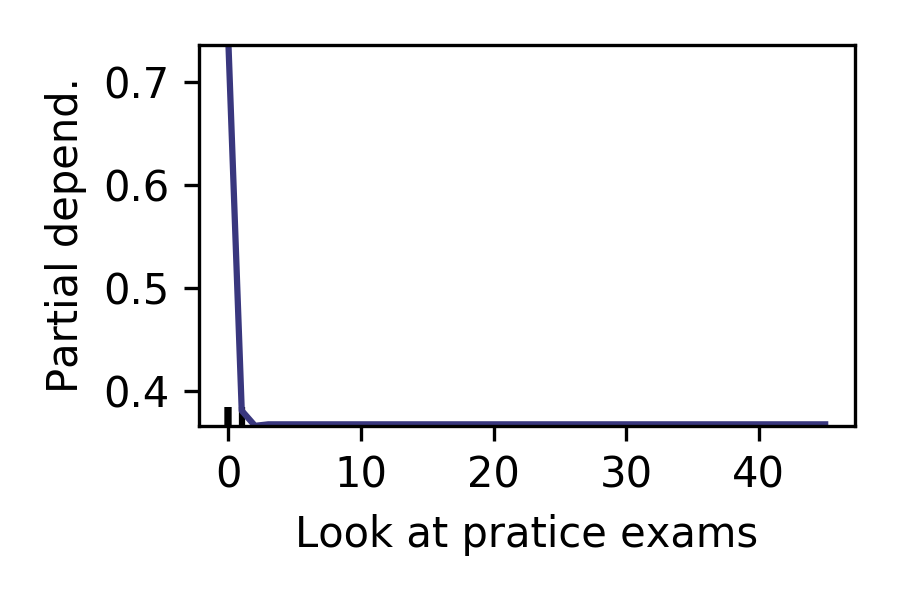} 
	\includegraphics[scale=.52]{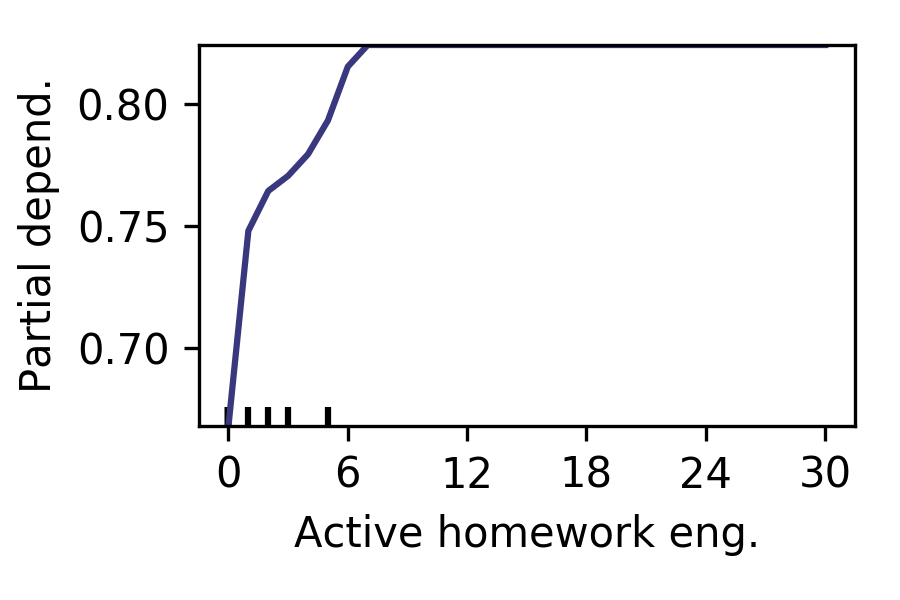} 
	\includegraphics[scale=.52]{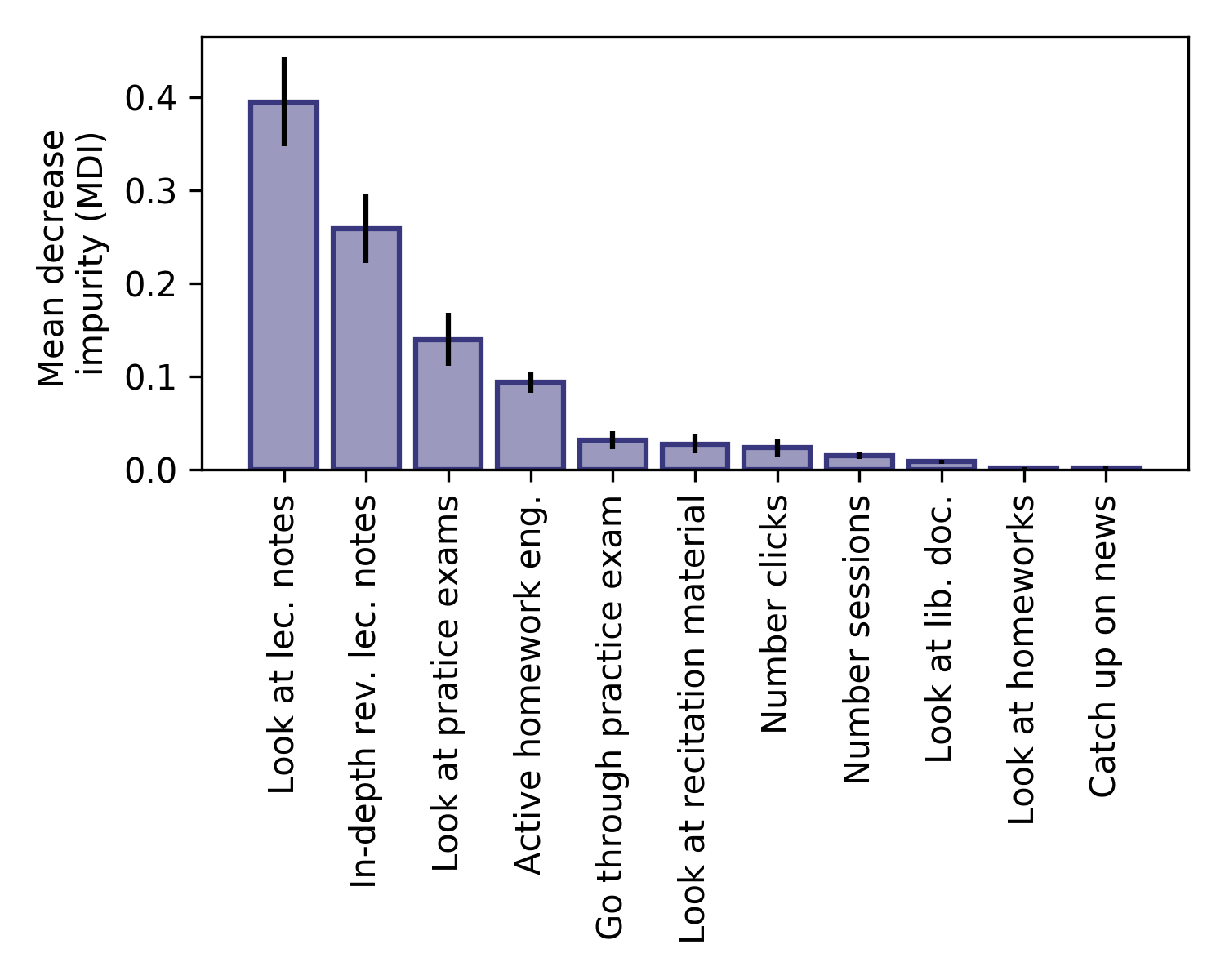}
	\vspace{-3ex}
	\caption{Relative feature importance for assessment week random forest prediction (1 = homework, 0 = exam) along with 95\,\% confidence intervals (bottom) and partial dependence plots for the most important features (top). Features include strategy counts from the clustered $n$-gram method, the number of clicks and the number of session clickstreams.}
	\label{fig:res}
\end{figure}

We extract strategy features for assessment week level prediction models by matching session clickstreams against the extracted frequent patterns. The results are summed up for each student-week combination and thus roughly represent how often a given student has used a strategy in a given assessment week.
After this aggregation, 91.03\,\% show at least one occurence of one of the patterns. %
We generally expect not all student click behavior to follow the extracted strategies or stringent strategies at all. Thus, it is unsurprising that %
some of the student week combinations do not involve any of the patterns.

We train a random forest classifier to predict the assessment type on pattern counts, the number of clicks and the number of sessions within a given week. 
A total of 80\,\% of the data is used for hyperparameter tuning and training, while 20\,\% is withheld for testing.
The model reaches a classification accuracy of 93.68\,\% on the training data which constitutes an evident improvement over the naive majority class prediction (71.90\,\% of the training data have the label homework).
Based on a permutation test, we find that the model performs better than random on the training set ($p<0.01$). A total of 100 permutations of labels were used for this evaluation.
Accuracy on the test set is 93.91\,\% which suggests sufficient generalization ability of the prediction model. 

The prediction model results suggest that students use the educational support system differently and employ the different strategies at different rates when preparing for exams as compared to doing homeworks. We examine feature importance in the model in order to gain more insights into these differences.
\Cref{fig:res} depicts the mean decrease in impurity (MDI) for splits at the different covariates, as well as partial dependence of the predictions on the most important features.
We see that predictions are mainly driven by pattern counts of the strategies look at lecture notes (MDI = 0.395), in-depth review of lecture notes (MDI = 0.259), look at practice exams (MDI = 0.140), and active homework engagement (MDI = 0.094).
Partial dependence plots show that while increased counts in the strategies related to lecture notes and practice exam engagement increase the probability that the model predicts an exam week, higher counts in the active homework engagement strategy increase the models likelihood of predicting an upcoming homework deadline. 
These results suggest that students approach to learning is driven by the kind of performance assessment they are given. It appears that the increased activity in exam weeks (see \Cref{fig:data}) is largely based on increased engagement with lecture notes and practice exams, while interactions with the homework related content is generally less pronounced.

\subsection{RQ3 Are student strategies indicative of grade outcomes?}

\begin{figure}[th]
	\centering
	\includegraphics[scale=.52]{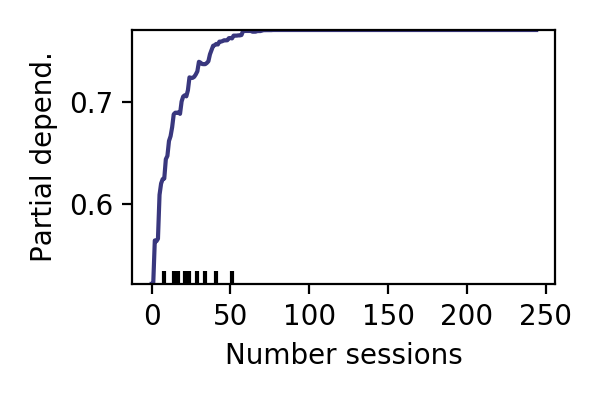} 
	\includegraphics[scale=.52]{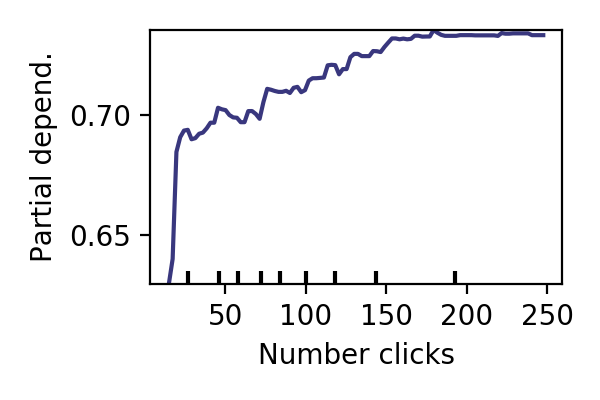} \\
	\includegraphics[scale=.52]{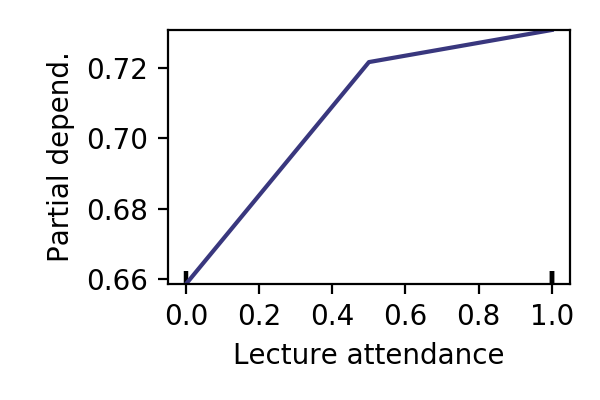} 
	\includegraphics[scale=.52]{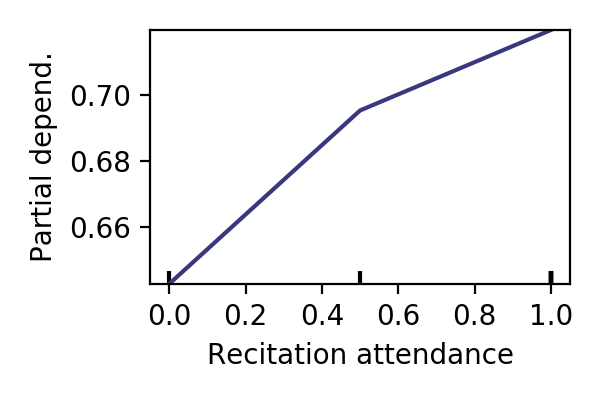} \\
	\includegraphics[scale=.52]{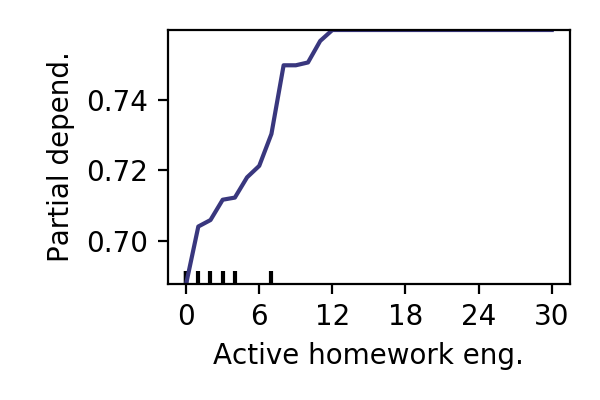} 
	\includegraphics[scale=.52]{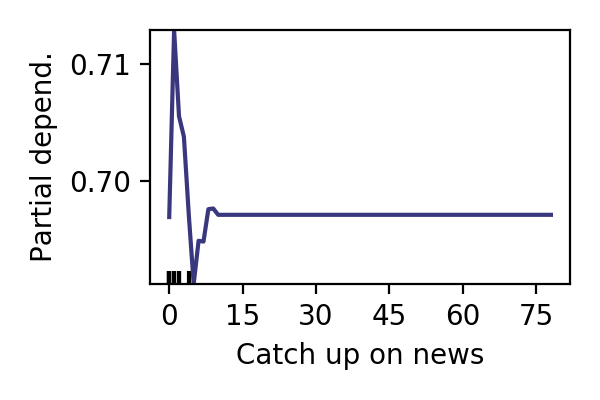} 
	\includegraphics[scale=.52]{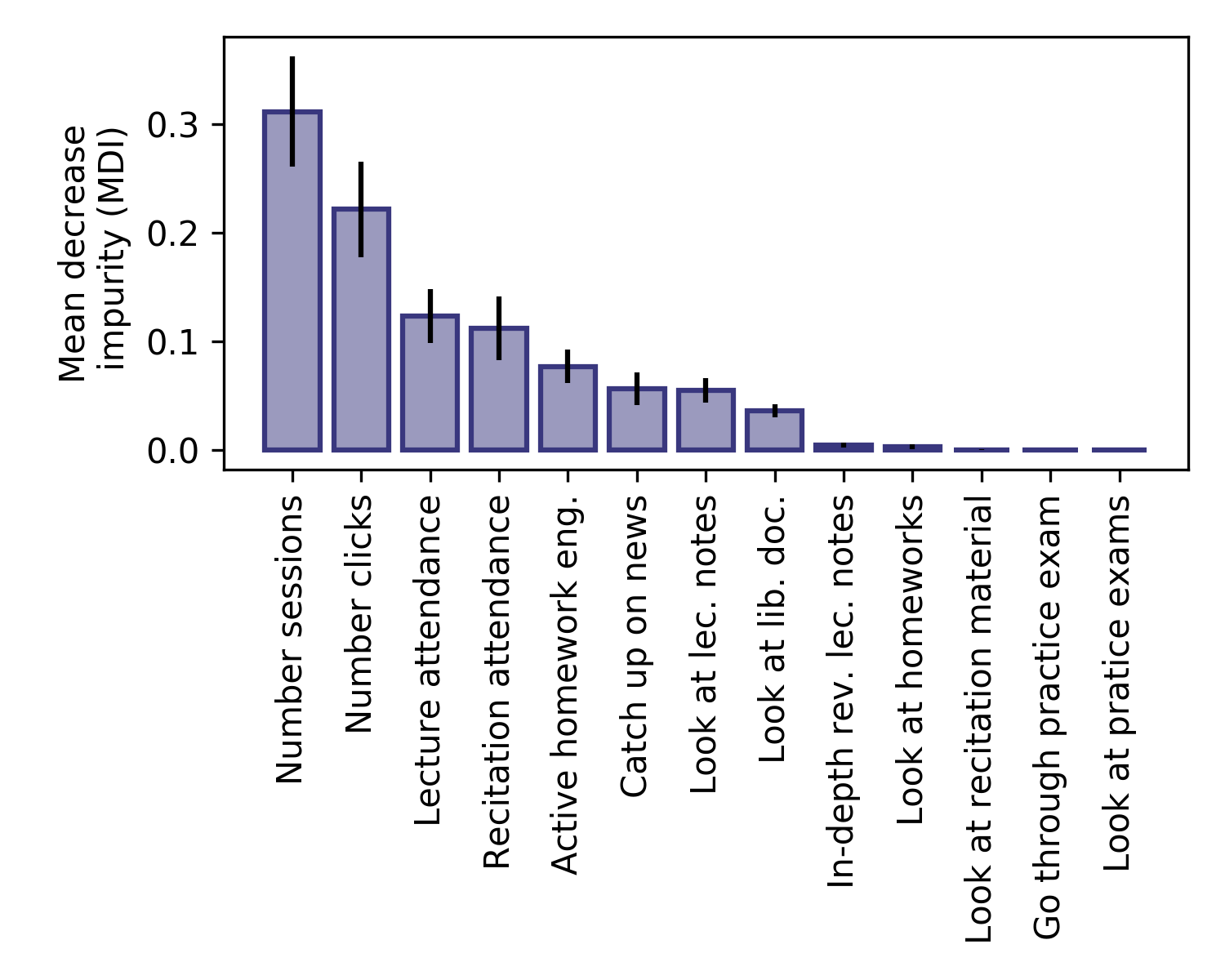}
	\vspace{-3ex}
	\caption{Relative feature importance for homework grade random forest prediction along with 95\,\% confidence intervals (bottom) and partial dependence plots for the most important features (top). Features include strategy counts from the clustered $n$-gram method, the number of clicks and session clickstreams, and attendance information.}
	\label{fig:res2}
\end{figure}

We train a random forest regression model to predict homework grades on a individual week and student level.
Features include students' strategy counts, the number of clicks, the number of sessions, and the mean attendance in both lectures and recitations.
Training is conducted on 80\,\% of available data while 20\,\% are withheld for testing. 
After hyperparameter tuning with 5-fold cross validation, the prediction model realizes a MSE of 0.046 on the training data set. 
A permutation test based on 100 permutations of labels shows a significant improvement over random performance with this model ($p<0.01$).
On the test set, the model attains a prediction MSE of 0.054 which suggests sufficient generalization ability.

\Cref{fig:res2} explores the importance of the different features for predictions and displays partial dependence relations for the most important covariates. Since we use MSE as impurity measure, the mean decrease in impurity (MDI) for a given feature effectively corresponds to the mean decrease in variance we receive by splitting at the feature.
We see that, in fact, the most relevant features appear to be the number of clickstream sessions (MDI = 0.311), the number of clicks (MDI = 0.221), lecture attendance (MDI = 0.123), and recitation attendance (MDI = 0.112).
Partial dependence plots reveal that increases in any of the above features increase the predicted homework score percentage by a relatively large margin of up to 20 percentage points.
Conversely, strategy counts appear to be less relevant for grade predictions with some exceptions.
Most notably, the predicted grade rises with the number of times students actively engaged in homeworks (MDI = 0.077).

Overall, our results show some success in prediction of homework grade outcomes.
The extracted features, including some of the pattern counts, add valuable information to the prediction model.
In particular, students who come back to Diderot more often and thus use an increased number of study sessions to solve their homeworks, and students who generally interact with the system at high rates are predicted to have better grade outcomes. In addition to time at task, the mere attendance in lectures and recitations increases students' grade outcome predictions. 
In fact, students in the our data set who attended at least one lecture in a given assessment week on average received a homework percentage grade of 76.58\,\%, while students who skipped lectures on average scored 55.86\,\%. %
For recitation attendance this corresponds to 74.34\,\% and 49.20\,\% respectively. 

Both of the discussed prediction models provide valuable insights for instructors and educational system design.
The tree-based ensemble methods are particularly suitable for initial modeling and processing of features on different scales. Their main advantage over many other models is the relatively straightforward explainability of predictions given partial dependence plots and measures of feature importance which renders them a useful approach to high stakes at-risk prediction.

\section{Conclusions}
\label{sec:conclusion}

Data from educational software systems provides insights into students' study behaviors.
While performance prediction in MOOCs has been explored extensively, similar studies for blended courses are scarce and often lack a deeper understanding of the underlying student strategies.
Based on fine-grained contextualizable click data collected through the non-commercial course support system Diderot, we explore how students interact with educational software systems, which strategies they employ to engage with course materials and in which ways strategies depend on the assessment type and drive performance.
Our contributions are two-fold: (1) We gain relevant understanding of students' learning behavior that both confirms and adds to the existing literature. (2) We propose new NLP-inspired approaches to analyzing student strategies' based on clickstream data in blended learning scenarios which typically come with moderately sized data sets.

On the educational side, our results provide valuable insights into how students interact with course systems. In line with previous research \cite{park_detecting_2017, agnihotri_mining_nodate}, we observe increased activity before deadlines, and, in particular, in the days leading up to an exam. Exam preparation appears to come with increased review of lecture notes as compared to homework solving.
In general, students seem to ask more questions related to homeworks as compared to other class materials such as lecture notes, recitation materials or practice exams.
At the same time, interactions with already existing posts such as liking or commenting seems to concentrate mostly on course-wide announcements, social posts and course feedback discussions and appears to be less common for direct questions on course materials.
Many of the derived features have some predictive power for performance outcomes. 
In particular, the number of study sessions, the number of clicks, attendance in lecture and recitation, and engaging with homework related course content are strong predictors for homework grades in our model.
The described observations are entirely based on data from a seven week period of a large sophomore level college course since technical difficulties prohibited collection of data for the remainder of the semester. In the future, more complete data (e.g. from an entire course, or even multiple courses such as the same course offering over several years) could provide an enhanced understanding of student behavior and allow the tackling of more complex problems such as the simultaneous prediction of homework and exam grades which, such as in our data, can have very different distributions.

The methods proposed in this work promise to be useful to a broad range of researchers and practitioners who find themselves analyzing activity log-data from blended courses, or are at the initial stages of developing early warning systems.
The key insight of this work is that hybrid NLP methods can be used to thoroughly analyze contexts of actions as well as frequent strategies in the relatively low-data setting of blended courses.
To the best of our knowledge, similar models have previously only been employed in the setting of MOOCs \cite[e.g.][]{wen_identifying_2014, pardos_analysis_2017}.
In fact, our analysis shows that topic models such as latent Dirichlet allocation can recover almost the same student strategies as more traditional data mining based pipelines of pattern extraction, and small versions of skip-gram neural networks can provide valuable insights into the context of student actions even with moderately sized data sets.

\printbibliography 

\end{document}